\documentclass[aps,twocolumn,showpacs,prl,amsmath,amssymb,floatfix,superscriptaddress]{revtex4-1}
\usepackage[utf8]{inputenc}
\usepackage{amsmath}
\usepackage{amssymb}
\usepackage{braket}
\usepackage{comment}
\usepackage{bbold}
\usepackage{bm}
\usepackage{mathbbol}
\usepackage{graphicx,color}
\usepackage{ulem} 
\usepackage{cancel}
\DeclareMathOperator{\Tr}{Tr}

\def\be{\begin{equation}}
\def\ee#1{\label{#1}\end{equation}}

\def\b{\beta}
\def\g{\gamma}
\def\G{\Gamma}
\def\d{\delta}
\def\D{\Delta}
\def\e{\epsilon}

\def\Th{\Theta}

\def\l{\lambda}
\def\La{\Lambda}

\def\r{\rho}
\def\s{\sigma}

\def\o{\omega}
\def\O{\Omega}

\def\i{\int}
\def\bGa{\mathbf \G}
\def\bga{\mbox{\boldmath$\g$}}

\def\bq{{\mathbf q}}
\def\bp{{\mathbf p}}

\def\BB#1{\bm{(}#1\bm{)}}

\begin{document}

\title{Quasistatic work processes: When slowness implies certainty}
\author{Juyeon Yi}
\affiliation{Department of Physics, Pusan National University, Busan 46241, Republic of Korea}
\affiliation{Department of Physics, Boston University, Boston, Massachusetts 02215, USA}
\author{Peter Talkner}
\affiliation{Department of Physics, University of Augsburg, D 86135 Augsburg, Germany}
\date{\today}

\begin{abstract} 
Two approaches are outlined  to characterize the fluctuation behavior of work applied to a system by a slow change of a parameter.  One approach uses the adiabatic theorems of quantum and classical mechanics, and the other one is based on the behavior of the correlations of the generalized coordinate that is conjugate to the  changed parameter.  Criteria are obtained under which the work done on small thermally isolated as well as on open systems ceases to fluctuate in a quasistatic process.      
\end{abstract}
\maketitle
The quasistatic change of a system parameter, like  of a system's volume or of an external electrical or magnetic field, provides one of the main building blocks of the theory of thermodynamic processes~\cite{Callen}. Such a quasistatic process can be performed in different ways, for example in thermal isolation or in contact with a heat bath. In the first case, the process is adiabatic, in the second one isothermal and in both cases it is reversible. Likewise, in both cases, work is performed on the system. This work is defined as the change of the energy of the total system consisting of the proper system and, if present, its environment with which it may exchange energy~\cite{Jarzynski2004,Campisi2009,CHT11}. For a macroscopic system in thermodynamic equilibrium, the work done in such a process is an extensive quantity and, hence, its fluctuations vanish in the thermodynamic limit~\cite{Kubo}. The amount of work performed on the system  then agrees with the change of the internal energy for an adiabatic process, and with the change of the free-energy for an isothermal process.  In the latter case, additionally the Jarzynski equality holds. It states that the expectation value of the negative exponentiated work per thermal energy is equal to the negative exponentiated free-energy change also taken per thermal energy~\cite{Jarzynski1997,Jarzynski2004,Campisi2009,J20}.  The simultaneous validity of the Jarzynski equation and the equality of work and free-energy change verifies the above mentioned fact that  the work does not fluctuate in macroscopic thermodynamic processes~\cite{vanishing}. 

In the present Letter we formulate conditions under which the work performed on a {\it microscopic} system in a quasistatic process does not fluctuate even though it can no longer be considered as an extensive quantity of a macroscopic system. These conditions cover and further specify the known cases described by classical Langevin equations~\cite{Hoppenau} and Markovian quantum mechanical master equations~\cite{Miller,Scandi}.  In the subsequent  analysis of quasistatic processes we follow two complementary routes, one based on the classical and quantum adiabatic theorems and the other one on the longtime behavior of a particular correlation function. While the conditions required by the first route often are difficult to control and also the important question how slow a particular process has to be in order to qualify as a quasistatic process, often is difficult to answer,  the second route provides a criterion of its validity as well as an access to the relevant timescales. The knowledge of these system-specific timescales may be used to design force protocols of finite duration with a minimal amount of work fluctuations.    

{\it Adiabatic invariance --} In general,  the amount of work applied to a system by a prescribed force protocol of finite duration varies with repeated realizations. The resulting work may also fluctuate for quasistatic parameter changes in isolated systems depending on their initial states as it is known from several examples~\cite{J2,Deffner,Yi11,Yi12}. On the other hand, the absence of work fluctuations is observed for systems undergoing quasistatic processes during which a sequence of stable and ergodic systems is realized provided that the process starts with the system  in a microcanonical initial state, i.e., with a well defined energy~\cite{Deng}. For classical systems, this follows from the adiabatic invariance of the phase-space volume~\cite{Hertz} and likewise for quantum systems from the quantum adiabatic theorem~\cite{Born} as will be demonstrated in more detail below. 

For a stable quantum system~\cite{stableQ,Ruelle}, the total number of states $\O(E,\l) = \Tr \Th \BB{ E- H(\l)} $ below any energy  $E$ is finite and  invariant under quasistatic changes of the control parameter $\l$ if the degrees of degeneracy of the eigenenergies of the Hamiltonian $H(\l)$ do not change upon a variation of the parameter~\cite{Born,Kato}.  Here, $\Th(x)$ denotes the Heaviside step function. For classical systems the according quantity is given by the volume $\O(E,\l)  = \i d\G \Th \BB{ E- H(\bq,\bp,\l) }$ of the phase-space below the energy $E$, where, $H(\bq,\bp, \l)$ denotes the Hamiltonian function depending on the positions $\bq$, momenta $\bp$ and the control parameter $\l$; further $d \G$ specifies a dimensionless infinitesimal phase-space volume element which is chosen such that the integral $\i d\G$ formally results as the classical limit of the trace $\Tr$. 

As a consequence of the stability of the system the volume $\O(E,\l)$ is a finite, monotonically increasing function of $E$, i.e., $\O(E_1,\l) \geq \O(E_2,\l)$ for $E_1 > E_2$, which is bounded from below by $\O(E,\l) =0$ for all $E$'s less than the ground state energy. Under the stated assumption of ergodicity for classical systems and the requirement that the degrees of degeneracy of the energy eigenvalues are constant throughout the protocol for quantum systems, the volume of phase-space, as well as the number of states, remain unchanged by a quasistatic process starting at $\l_i$ with the energy $E_i$ and ending at $\l_f$ with the energy $E_f$~\cite{Hertz,Born}, i.e.,
\be
\O(E_i,\l_i) = \O(E_f,\l_f)\:. 
\ee{OO}
Hence, also the microcanonical entropy $S(E,\l) = k_B \ln \O(E,\l) $~\cite{Gibbs,Hertz,HHD} remains constant in a quasistatic work process, i.e.,
\be
S(E_i,\l_i) = S(E_f,\l_f)\:.
\ee{S} 
With the monotonicity of $\O(E,\l)$ as a function of $E$ the Eq.~(\ref{OO}) implies the uniqueness of the final energy $E_f$ which can be expressed in terms of the inverse function $\O^{-1}$ as 
\be
E_f(\l_i,\l_f,E_i) = \O^{-1}\BB{ \O(E_i,\l_i),\l_f }\:.
\ee{Ef}
Accordingly, the work results as a unique function of the initial energy, the initial parameter value $\l_i$ and the final one, $\l_f$ given by
\be
w(\l_i,\l_f,E_i) = E_f(\l_i,\l_f,E_i) - E_i\:.
\ee{w}   
Because in most cases, the work will depend on the initial energy~\cite{ho}, one might expect that quasistatic processes starting in other than a  microcanonical state, namely with an initial energy distributed with a non-vanishing variance are characterized by a more or less broad distribution of work having a non-vanishing  variance of the work, in accordance with the above cited works~\cite{J2,Deffner,Yi11,Yi12} all of which are dealing with closed low-dimensional systems. 

In order to better understand the behavior of quasistatic processes in the large class of open systems, without specific assumptions being made about their dynamics, we enlarge the considered system by taking into account all of its environment until a total, thermally isolated closed system is obtained.    
Then the Hamiltonian of the total system $H_{\text{tot}}(\l)$ is given by the sum 
of the Hamiltonians of the bare system $H_s(\l)$, the bare bath $H_B$ and an interaction term $H_I$, yielding
\be
H_{\text{tot}}(\l) =H_S(\l) + H_B + H_I\:.
\ee{Htot}
Here, the controllable parameter $\l$ exclusively enters the system Hamiltonian $H_S(\l)$. The change in the energy of the total system due to a variation of $\l$ can then be identified with the work performed on the system~\cite{Campisi2009,CHT11}. We note that the interaction part needs not be small compared to the system Hamiltonian. However, we do assume that the environment is much bigger than the system and that {the system plus environment can be considered as a normal system~
\cite{Kubo,normal}. As such the energy and also the entropy of the total system are extensive with respect to the number $N$ of the constituents of the bath.
Because the work is of the order $N^0$ compared to the initial and final energies which are of the order $N$, the entropy at the final energy $E_f = E_i +w(\l_i,\l_f,E_i)$ and parameter value $\l_f$ can be expanded about the initial energy yielding the following expression for the work
\be
w(\l_i,\l_f,E_i) =T(E_i,\l_f) \big (S(E_i,\l_i) - S(E_i,\l_f) \big )\:
\ee{wSS}  
with corrections being of the order $N^{-1}$.  
For a small system in contact with a large environment the difference of entropies with the same energy at different parameter values is indeed an intensive quantity as follows from the partial derivative of the entropy with respect to the parameter which can be expressed as $\partial S(E,\l)/\partial \l = T^{-1}(E,\l) \langle \partial H_S(\l)/\partial \l \rangle_E$, where $\langle \bullet \rangle_E = \o^{-1}(E,\l) \Tr \bullet \delta\BB{E-H_{\text{tot}}(\l)}$ denotes a microcanonical average and   $T(E,\l) =(\partial S(E,\l)/\partial E )^{-1}$ is the microcanonical temperature, which is also intensive. The parameter-dependence of the temperature is negligible because of  $\partial T(E,\l)/\partial \l = \mathcal{O}(N^{-1})$~\cite{Tl}. Hence, $T(E,\l) = T(E)$.  
Under the assumption that the total system is normal, the microcanonical and the canonical ensembles are equivalent. Using $T(E) S(E,\l) = E - F(T(E),\l)$ we obtain for the work \cite{Campisi2009,CHT11,TH20} 
\be
\begin{split}
w(\l_i,\l_f,E_i) &= F_{\text{tot}}\BB{T(E_i),\l_f} -F_{\text{tot}}\BB{T(E_i),\l_i}\\
&= F_S\BB{T(E_i),\l_f} - F_S\BB{T(E_i),\l_i}\:,
\end{split}
\ee{wc}
where the free-energy of the open system is determined by the difference of the free-energies of the total system and the bare bath, 
$F_S(t,\l) = F_{\text{tot}}(T,\l) - F_B(T)$.  
Most remarkably, the work depends only via the microcanonical temperature on the initial energy. 
Because in the thermodynamic limit the microcanonical and the canonical temperatures agree with each other~\cite{Ruelle}, also for a small system the work of an isothermal process does not fluctuate and consequently coincides with the difference of the free-energies of the considered open system at the end and the beginning of the protocol.  

Even though this result is formally exact, for most experimental situations it will be difficult to determine whether the conditions for the adiabatic theorems are properly satisfied, in particular, because for large systems the energy spectra become very dense and hence level crossings upon changes of $\l$ may become quite likely. Even if no crossings occur, the timescale for which the theorems apply may be unreasonably large. This might happen for classical and for quantum systems. In the rest of this Letter we provide an alternative approach which also yields information about the required duration of a protocol to qualify as quasistatic.

{\it Correlation expression of the work variance --} In order to prove that the work does not fluctuate, it is sufficient to show that the variance of the work, $\s^2_w = \langle w^2 \rangle - \langle w \rangle^2$, vanishes. The moments of the work follow from the characteristic function $G(u)$ as derivatives at $u=0$, $\langle w^n \rangle = (-i)^n d^n G(u)/ d u^n |_{u=0}$~\cite{Lukacs}. For a quantum system, governed by the Hamiltonian $H\BB{\l(t)}$ the characteristic function is given by, see~\cite{TLH,CHT11},  
\be
G(u) = \Tr e^{i u H^H(\l_f,t_f)}e^{-i u H(\l_0)}\bar{\r}(0)\:,
\ee{G} 
where $H^H\BB{\l(t),t} = U^\dagger_{t,o} H\BB{\l(t)} U_{t,o} $ denotes the Hamiltonian in the Heisenberg picture at time $t\geq 0$ with the time-evolution operator $U_{t,s}$ governed by the Schr\"odinger equation $i \hbar \partial U_{t,s} /\partial t = H\BB{\l(t)}  U_{t,s}$ satisfying $U_{s,s} = \mathbb{1}$.  The density matrix $\bar{\r}(0) = \sum_n \Pi_n \r(0) \Pi_n$ is the diagonal part of the actual initial density matrix $\r(0)$ with respect to the energy eigen-basis of the Hamiltonian at the initial time $t=0$ as specified by the projection operators $\Pi_n$ onto  the energy eigenstates having the energy eigenvalues $E_n$.
The same formal expression holds for a classical system if one interprets the trace as the integral over the system's phase-space and the Hamilton operators as the corresponding Hamiltonian functions. Furthermore, the Hamiltonian operator in the Heisenberg picture corresponds to the Hamiltonian function taken at the phase-space trajectory starting at a phase-space point on which the integration is performed and being evaluated till the end of the force protocol. The initial density matrix is replaced by the initial phase-space probability density function (pdf),  whereby the map $\r \to \bar{\r}$ is ineffective for classical systems. The first two moments of the work can be expressed in terms of the difference of the Hamiltonians according to
\be
\langle w^n \rangle = \langle \left [ H^H(\l_f,t_f) - H(\l_0) \right ]^n \rangle\:,\quad n=1,2
\ee{wn}  
where the average is performed with respect to the projected initial state $\bar{\r}(0)$. For classical systems this expression holds for all $n$, for quantum systems it does so only if the two Hamiltonians commute with each other. In the general quantum case the validity of Eq. (\ref{wn}) as indicated is restricted to the first two moments~\cite{EN}.  Writing the difference of the Hamiltonians as an integral of the total derivative of the timedependent Hamiltonian, one obtains with $d H^H\BB {\l(t),t }/d t =   \partial H^H\BB {\l(t),t }/\partial t = \dot{\l}(t) Q^H(t)$ expressions for the first two moments of work in terms of the generalized coordinate $Q(t) = \partial H(\l)/\partial \l |_{\l = \l(t)}$, which is conjugate to the parameter $\l$~\cite{lambda}. As for the Hamiltonian the upper index $H$ indicates the Heisenberg picture. The average work then becomes
\be
\langle w \rangle = \i_0^{t_f} dt \dot{\l}(t) \langle Q^H(t) \rangle
\ee{wQ}
and the variance
\be
\begin{split}
\s^2_w &= \i_0^{t_f} dt_1  \i_0^{t_f} dt_2 \dot{\l}(t_1) \dot{\l}(t_2) \langle \d Q^H(t_1) \d Q^H(t_2) \rangle\\
&= 2\i_0^{t_f} dt \i_0^t ds \dot{\l}(t) \dot{\l}(t-s) C(t,s)\:,
\end{split}
\ee{w2Q}
where $\d Q^H(t) = Q^H(t) -\langle Q^H(t) \rangle$ denotes the fluctuation and $C(t,s) =\big (\langle \d Q^H(t) \d Q^H(t-s) \rangle +  \langle \d Q^H(t-s) \d Q^H(t) \rangle \big )/2$  denotes the symmetrized autocovariance of the generalized coordinate. All averages are again taken with respect to the initial projected density matrix $\bar{\r}(0)$ or the respective classical pdf. 
These expressions for the first two moments are valid for quantum and classical systems and for any protocol. For classical systems, the Heisenberg picture of the generalized coordinate has to be replaced by $Q\BB{\bGa(t,\bga),\l(t)}$ with $Q(\bga,\l) =\partial H(\bga,\l)/\partial \l$ and $\bGa(t,\bga)$ being the solution of the Hamiltonian equations of motion starting at the phase-space point   $\bGa(0,\bga) =\bga$. 

{\it Quasistatic protocols --} Any quasistatic protocol $\La_{\text{qs}}$ can be obtained from
a conveniently chosen protocol $\La=\{\l(t)|0\leq t \leq t_f \}$  of finite duration $t_f$, performed at a finite speed, by means of a uniform scaling of the time. The quasistatic protocol is then obtained by performing the limit $\e \to 0$ of the corresponding $\e$-protocol $\l_\e(t) = \l(\e t)$. It can be formally expressed as $\La_{\text{qs}} = \lim_{\e \to 0} \{\l_\e(t)|0\leq t \leq t_f/\e \}$.
Under the assumption that the considered system is stable for all parameter values during the protocol, the average generalized coordinate will be a bounded function of $t$ for all values of $\e$ and hence the average work will approach a finite value for all values of $\e$ including the quasistatic limit leading to a bound of the average work given by $|\langle w \rangle | \leq \ell \max_{t\in [0, t_f/\e]} |\langle Q^H(t) \rangle|$, where $\ell= \i_0^{t_f} dt |\dot{\l}(t)|$. Likewise, the variance is bounded with $\s^2_w \leq \ell^2  \max_{t\in [0, t_f/\e]} \langle \big (\d Q^H(t) \big )^2 \rangle$. The variance of work becomes in the quasistatic limit
\be
\s^2_w = 2\lim_{\e \to 0} \e^2 \i_0^{t_f/\e} dt \i_0^t ds \dot{\l}(\e t) \dot{\l}(\e(t-s)) C(t,s)\:.
\ee{swqs}
This expression can be further analyzed under the assumption that  the fluctuations of the generalized coordinate  become uncorrelated at largely separated times, i.e., that the symmetrized autocovariance $C(t,s)$ vanishes with increasing separation $s$. Then the integration in Eq.~(\ref{swqs}) can be split into a triangular part up to a characteristic time $t_c$ beyond which the autocovariance becomes negligible, $C(t,s) \approx 0$ for $s>t_c$,   and a trapezoid part, i.e., symbolically $\i_0^{t_f/\e}dt \i_0^t ds = \i_0^{t_c}dt \i_0^t ds + \i_{t_c}^{t_f/\e}dt \i_0^t ds$. The triangular part yields a finite contribution to the integral on the right hand side of Eq.~(\ref{swqs}) which, together  with the $\e^2$ factor, results in a vanishing contribution to the work variance in the quasistatic limit.  Taking into account the vanishing of $C(t,s)$ for $s>t_c$, the trapezoid domain can be reduced to a rectangular one, $\i_{t_c}^{t_f/\e} dt \i_0^t ds \to  \i_{t_c}^{t_f/\e} dt \i_0^{t_c} ds$. In this integral, the $s$-dependence of $\dot{\l}\BB{ \e(t-s)}$ can be neglected. The remaining $s$-integral can be expressed as $\i_0^{t_c} ds C(t,s) =\vartheta(t) \langle \big (\d Q(t) \big )^2 \rangle$ where $\vartheta(t)$ is the correlation time \cite{Papoulis} which is of the same order of magnitude as the characteristic decay time $t_c$.  For a protocol visiting only stable systems, both  $\vartheta(t)$ and the variance of the generalized coordinate $\langle \big (\d Q(t) \big )^2 \rangle$ are bounded functions of the time. Therefore the $t$ integral is proportional to $t_f/\e$ and, finally, in combination with the $\e^2$ factor  the work variance vanishes in the quasistatic limit proportionally to $\e$. This fact is also of practical importance because it opens the possibility to find a compromise of a tolerable noise level of the work at a finite protocol duration.          

The vanishing of the autocovariance of the generalized coordinate is known as mixing and requires a sufficiently irregular dynamical behavior of the considered system as it can be expected for an open system as a small part of a much larger system. However, also for closed classical chaotic systems~\cite{LL} their ensuing ergodic and mixing behavior~\cite{AA}  entails the decorrelation of the fluctuations of the generalize coordinate at large time separations.  
For quantum systems a generally accepted definition of chaos is still lacking~\cite{H}. Also the fact that for systems with a finite dimensional Hilbert space the autocovariance of any observable is an almost-periodic function strictly speaking makes it impossible that it vanishes for all time lags beyond a particular time $t_c$. However, already for modestly large Hilbert space dimensions the almost recurrence time $t_P$ after which the equal-time value of the autocovariance is nearly recovered, typically is astronomically large, such that for any experimentally relevant situation with $t_c \ll t_f/\e \ll t_P$ the fluctuations of the work are considerably suppressed. Only if the duration of the process approaches the recurrence time $t_P$, the work fluctuates like it does for short experiments with $t_c > t_f/\e$. 
Even for the best possible isolation of a system, those interactions that can safely be neglected during a  long initial period may become effective on the scale of the recurrence time and most likely  will destroy those subtle coherences in the system that are necessary for a recurrence.       

For open systems at any interaction strengths between the system and its environment~\cite{TH20},  the presence of the environment will typically induce enough randomness into the dynamics of the proper system such that a decay of the generalized coordinate autocovariance can be expected independent of the particular nature of the open system's dynamics, whether it is Markovian or not. 
A rough estimate of the characteristic time $t_c$ can be obtained as the maximal
autocovariance decay time taken over the set of autonomous systems with frozen parameter values.           

{\it Conclusions --} Two complementary approaches are presented to explain the particular behavior of quasistatic  work processes of isolated as well as open systems. One approach is based on the adiabatic theorems of classical and quantum mechanics providing the conditions under which the work performed on the system does not fluctuate. While this approach is based on statistical mechanical notions characterizing the phase-space of the considered system, the second approach is focused on dynamical aspects of the generalized coordinate that is conjugate to the slowly varied parameter causing the forcing. Whenever the fluctuations of the generalized coordinate become uncorrelated on a  timescale that is much shorter than the duration of the forcing, the variance of the work vanishes.  Knowing this specific timescale opens the possibility to run work processes within finite time with only small fluctuations. For small engines this means that a high efficiency at finite power may be achieved.                 

{\it Acknowledgment} We thank Peter H\"anggi for numerous discussions and Anatoli Polkovnikov for his great help
and hospitality. This work was supported by the Finan-
cial Supporting Project of Long-term Overseas Dispatch
of PNU’s Tenure-track Faculty.


\begin{thebibliography}{99}
\bibitem{Callen} H.B. Callen, {\it Thermodynamics and an Introduction to Thermostatistics} (John Wiley, New York 1985).
\bibitem{Jarzynski2004} C. Jarzynski, {\it Nonequilibrium work theorem for a system strongly coupled to a thermal environment}, J. Stat. Mech. (2004) P09005.  
\bibitem{Campisi2009} M. Campisi, P. Talkner, and P. H\"anggi, {\it Fluctuation theorem for arbitrary open quantum systems}, Phys. Rev. Lett. {\bf 102}, 210401 (2009).
\bibitem{CHT11} M. Campisi, P. H\"anggi, and P. Talkner, {\it Colloquium: Quantum fluctuation relations: Foundations and applications}, Rev. Mod Phys. {\bf 83}, 771 (2011).

\bibitem{Kubo} R. Kubo, {\it Statistical Mechanics: An Advanced Course with Problems and Solutions}, North Holland, Amsterdam 1965.
\bibitem{Jarzynski1997} C. Jarzynski, {\it Nonequilibrium equality for free energy differences}, Phys. Rev. Lett. {\bf 78}, 2690 (1997). 
\bibitem{J20} C. Jarzynski, {\it Fluctuation relations and strong inequalities for thermally isolated systems}, Physica A {\bf 552}, 122077 (2020).
\bibitem{vanishing} The vanishing of work fluctuations under these  conditions  is based on the mathematical fact that a random number $x$, with $\langle x \rangle =0$ and $\langle e^{x}\rangle =1$ assumes  the only value $x=0$. Here, the random number $x$ in question is given by the difference of work and free energy change, both taken per thermal energy, $x = \b(w -\D F)$.         

\bibitem{Hoppenau} J. Hoppenau, and A. Engel, {\it On the work distribution in quasi-static processes}, J. Stat. Mech: Theory and Experiment, P06004 (2013).

\bibitem{Miller} H.J.D. Miller, M. Scandi, J. Anders, and M. Perarnau-Llobet, {\it Work fluctuations in slow processes: Quantum signatures and optimal control}, Phys. Rev. Lett. {\bf 123}, 230603 (2019).
\bibitem{Scandi} M. Scandi, H.J.D. Miller, J. Anders, and M. Perarnau-Llobet, {\it Quantum work statistics close to equilibrium}, Phys. Rev. Res. {\bf 2}, 023377 (2020). 
\bibitem{J2} C. Jarzynski, {\it Equilibrium free energy difference from nonequilibrium measurements: A master-equation approach}, Phys. Rev. E {\bf 56}, 5018  (1997).
\bibitem{Deffner} S. Deffner, O. Abah, and E. Lutz, {\it Quantum work statistics of linear and nonlinear parametric oscillators}, Chem. Phys. {\bf 375}, 200 (2010).
\bibitem{Yi11} J. Yi, and P. Talkner, {\it Work statistics of charged noninteracting fermions in slowly changing magnetic fields}, Phys. Rev. E {\bf 83}, 041119 (2011).
\bibitem{Yi12} J. Yi, Y.W. Kim, and P. Talkner, {\it Work fluctuations for Bose particles in grand canonical initial states}, Phys. Rev. E {\bf 85}, 051107 (2012).
\bibitem{Deng} J. Deng, A.M. Tan, P. H\"anggi and J. Gong, {\it Merits and qualms of work fluctuations in classical fluctuation theorems}, Phys, Rev. E {\bf 95}, 012106 (2017).
\bibitem{Hertz} P. Hertz, {\it \"Uber die mechanischen Grundlagen der Thermodynamik}, Ann. Phys. (Leipzig) {\bf 33}, 537 (1910). 
\bibitem{Born} M. Born, and V.A. Fock, {\it Beweis des Adiabatensatzes}, Z. Phys. {\bf 51}, 165 (1928).
\bibitem{stableQ} A stable system is confined in space. Its energy is bounded from below.  For a quantum system the eigen-energies then form a discrete spectrum which accordingly is bounded from below~\cite{Ruelle}.  For classical systems, all subsets $\mathcal{G}_E=\{\bq,\bp|H(\bq,\bp) \leq E \}$ of the phase-space with a given maximal energy $E$ are bounded and hence possess a finite volume. For unstable quantum systems the energy spectrum is unbounded from below or it possess continuous parts and accumulation points, properties all leading to diverging numbers of states below a given energy. For classical systems the subsets $\mathcal{G}_E$ are unbounded for some or possibly all energies $E$.       
\bibitem{Ruelle} D. Ruelle, {\it Statistical Mechanics}, W. A. Benjamin Inc., Amsterdam, 1969.
\bibitem{Kato} T. Kato, {\it On the adiabatic theorem of quantum mechanics}, J. Phys. Soc. Jpn. {\bf 5}, 435 (1950).
\bibitem{Gibbs} J.W. Gibbs, {\it Elementary Principles in Statistical Mechanics}, Charles Scribner's Sons, New York, 1902.
\bibitem{HHD} S. Hilbert, P. H\"anggi, and J. Dunkel, {\it Thermodynamic laws in isolated systems}, Phys. Rev. E {\bf 90}, 062116 (2014). 
\bibitem{ho} An exception from this rule is provided by a harmonic oscillator which is linearly and quasistatically driven by an external force that couples to the coordinate~\cite{TBH}. The work applied to the oscillator in such a process  depends on the initial and the final parameter values but not on the initial energy.
\bibitem{TBH} P. Talkner, P.S. Burada, and P. H\"anggi, {\it Statistics of work performed on a forced harmonic quantum oscillator}, Phys. Rev. E {\bf 78}, 011115 (2008).
\bibitem{normal} 
Any system made of subsystems with local, i.e., short ranged, mutual interactions becomes a normal system in the thermodynamic limit~\cite{Kubo}. 
\bibitem{Tl} Using the definition of temperature one obtains $\partial T(E,\l)/\partial \l = -T^2(E,\l)\partial^2 S(E,\l)/(\partial E \partial\l)$. Because, as demonstrated above, $\partial S(E,\l)/\partial \l$ is intensive, its derivative with respect to $E$ is $\mathcal{O}(N^{-1})$.
\bibitem{TH20} P. Talkner. and P. H\"anggi, {\it Colloquium: Statistical mechanics and thermodynamics at strong coupling: Quantum and classical}, Rev. Mod. Phys. {\bf 92}, 041002 (2020).

\bibitem{Lukacs} E. Lukacs, {\it Characteristic Functions}, 2nd. ed., Griffin, London, 1970.Mod. Phys. {\bf 92}, 041002 (2020).   
\bibitem{TLH} P. Talkner, E. Lutz, and P. H\"anggi, {\it Fluctuation theorems: Work is not an observable}, Phys. Rev. E {\bf 75}, 050102(R) (2007).
\bibitem{EN} A. Engel, and R. Nolte, {\it Jarzynski equation for simple quantum systems: Comparing two definitions of work}, EPL {\bf 79}, 10003 (2007).
\bibitem{lambda} The generalization to parameters with more than a single component leading to a multi-component generalized coordinate is straightforward. 
\bibitem{Papoulis} A. Papoulis, {\it Probability, Random Variables, and Stochastic Processes}, McGraw-Hill, Aukland 1987.
\bibitem{LL} A.J. Lichtenberg, and M.A. Lieberman, {\it Regular and Chaotic Dynamics}, 2nd ed., Springer Verlag, Berlin 1983.
\bibitem{AA} V.I. Arnold, and A. Avez, {\it Ergodic Problems of Classical Mechanics}, W.A. Benjamin Inc. 1968.
\bibitem{H} F. Haake, {\it Quantum Signatures of Chaos},  2nd ed., Springer Verlag, Berlin 2001.






\end{thebibliography}
\end{document}